\documentclass[manuscript]{aastex}   	
\usepackage{natbib}
\providecommand{\e}[1]{\ensuremath{\times 10^{#1}}}
\usepackage{amssymb}
\usepackage{amsmath,amsfonts,amsthm,amssymb}

\begin{document}
\title{New Gapless COS G140L Mode Proposed for Background-Limited Far-UV Observations}

\author{Keith Redwine\altaffilmark{1}, Stephan R. McCandliss\altaffilmark{2}, and Wei Zheng\altaffilmark{3}}
\affil{Department of Physics and Astronomy, Johns Hopkins University,
	Baltimore MD 21218}

\author{Brian Fleming\altaffilmark{4} and Kevin France\altaffilmark{5}}
\affil{† Laboratory for Atmospheric and Space Physics, University of Colorado,
	Boulder CO 80309-0390}

\author{Steven Osterman\altaffilmark{6}}
\affil{Johns Hopkins University Applied Physics Laboratory, Laurel Md 20723}

\author{J. Chistopher Howk\altaffilmark{7}}
\affil{Department of Physics, University of Notre Dame,
	Notre Dame IN 46556}

\author{Scott F. Anderson\altaffilmark{8}}
\affil{Department of Physics, University of Washington, 3910 15th Ave. NE
Seattle WA 98195-1560}
\and
\author{Boris T. G{\"a}ensicke\altaffilmark{9}}
\affil{Department of Physics, University of Warwick, Gibbet Hill Road, Coventry CV4 7AL UK}

\altaffiltext{1}{kredwin1@jh.edu}
\altaffiltext{2}{stephan.mccandliss@jhu.edu}
\altaffiltext{3}{wzheng3@jhu.edu}
\altaffiltext{4}{brian.fleming@colorado.edu}
\altaffiltext{5}{kevin.france@colorado.edu}
\altaffiltext{6}{steve.osterman@jhuapl.edu}
\altaffiltext{7}{jhowk@nd.edu}
\altaffiltext{8}{anderson@astro.washington.edu}
\altaffiltext{9}{boris.gaensicke@warwick.ac.uk}

\begin{abstract}
Here we describe the observation and calibration procedure for a new G140L observing mode for the Cosmic Origins Spectrograph (COS) aboard the {\it Hubble Space Telescope} (HST).  This mode, CENWAV = 800, is designed to move the far-UV band fully onto the Segment A detector, allowing for more efficient observation and analysis by simplifying calibration management between the two channels, and reducing the astigmatism in this wavelength region.  We also describe some of the areas of scientific interest for which this new mode will be especially suited.
\end{abstract}

\section{Introduction}
The demonstrated sensitivity of COS~\citep{Green:2012,Osterman:2011} G140L mode to wavelengths below 1150 \AA\citep{McCandliss:2010} has opened a new window to the universe previously inaccessible to Hubble, and enabled a number of compelling science investigations, one of which is the search for ionizing hydrogen Lyman continuum (LyC) photons from starforming galaxies at low redshift in order to characterize the escape process ~\citep{Finkelstein:2014}. To take full advantage of the FUV capabilities of COS, we proposed and calibrated a new observing mode where the CENWAV setting, the nominal shortest wavelength to appear at the short wave end of the Segment A detector, was set to 800 \AA, providing contiguous wavelength coverage from 900 to 1850 \AA\, on the Segment A detector. This new mode has three significant advantages: 1) a contiguous spectral coverage on a single detector segment, which will simplify tracking of flux, wavelength and flat-field calibrations; 2) a $\sim$ 2 times lower background from reduced astigmatism at the short wavelength end where the effective area is lowest, which will boost the signal-to-noise of background limited observations; and 3) a more efficient use of observing time for programs requiring the full far-UV wavelength coverage by eliminating the need for a grating change.  

We describe the flux and wavelength calibration of the new mode, using the calibration target AV 243, a stable O6V star in the Small Magellanic Cloud. We observed AV 243 in Cycle 19, and also utilized archival spectra~\citep{Sahnow:2000} of AV 234 previously acquired by the {\it Far-Ultraviolet Spectroscopic Explorer} (FUSE)~\citep{FUSE_Moos:2000}, the Faint Object Spectrograph (FOS)~\citep{FOS_Greenfield:1991} once aboard HST, and the {\it International Ultraviolet Explorer} (IUE)~\citep{IUE_Boggess:1978}, which guided our analysis of the new CENWAV=800 mode of COS G140L. We determine the best focus position and characterize the astigmatism as a function of wavelength, as well as providing a wavelength calibration and effective area calculation.

\section{Data}
	In order to establish this new mode, the selected source needed to have a high density of relatively narrow atomic and molecular absorption lines to determine focus and a wavelength solution.  It was also required to be bright enough to supply a high enough signal-to-noise over the allotted orbit sweeps and be dim enough to not violate the bright object protection of COS~\citep{COShandbookcycle19}.  The integrated flux from AV 243 is known from previous observations by FUSE and FOS.  A global count rate of 11141.1 counts s$^{-1}$ was observed in the COS extraction window in this program, which was below the global limit of 15000 counts s$^{-1}$ but high enough to provide a good volume of data.

\begin{figure}
\begin{center}
\includegraphics[scale=0.75]{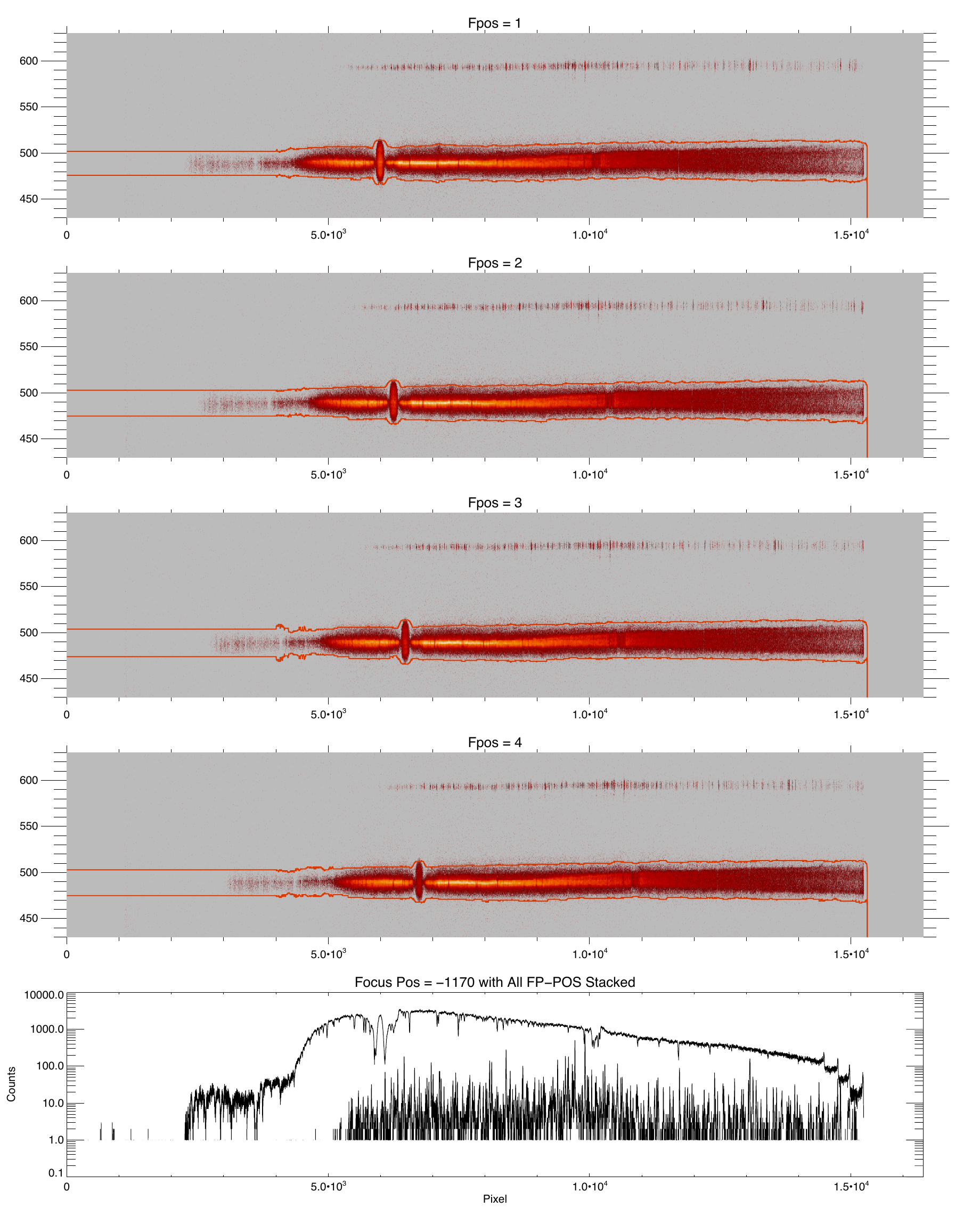}
\caption{\label{fpos1170} Calibrated spectra in pixel-space of four FP-POS positions, in Fpos=-1170 focus setting, with a one extracted one-dimensional spectrum of the on-target region and the PtNe lamp region of the detector.  The red lines in the two-dimensional windows show the extraction windows for calculation of the one-dimensional spectrum.}
\end{center}
\end{figure}

	We were granted time for three orbits of integration time in cycle 19 (HST Proposal ID \#12501) on the target.  Over the course of the observation, with a central wavelength setting rotated to 800 \AA, the spectrum was moved to the four different FO-POS positions.  Each of the focus settings used, Fpos = -370, -770, -870, -970, -1070, -1170, moved the focus of the spectrum across the detector in the dispersion direction.  Note that these focus shifts given by Fpos are distinct from the FP-POS settings; in a typical COS observation the FP-POS will cycle through each setting, shifting the entire spectrum in the dispersion direction in order to preserve consistent sensitivity across the detector.  The Fpos settings are not generally adjusted over an observation, and shift the position of the most sharply focused point of the spectrum, while not moving the spectrum itself.  An example of the raw images and resulting spectrum is shown in Figure~\ref{fpos1170}.

One-dimensional spectra were extracted, using an optimal method for determining a summation envelope following the astigmatism.  For a column containing $N$ counts on the detector, we set symmetric bounds around the centroid of the column such that the bounds exclude a proportion of $\sqrt{\frac{1}{N}}$ of the data.  The counts are then simply totaled for each column, resulting in a one-dimensional spectrum.  In Figure~\ref{fpos1170}, these bounds are explicitly drawn onto four of the two-dimensional COS spectra, along with the resulting one-dimensional extraction for the best focus position.  For columns with lower count rates the extraction region is larger, exhibited by the larger regions contained by the bounds of the extraction regions of the spectra.  This procedure assumes a symmetric cross-dispersion profile, which isn't strictly the case, but the error due to asymmetry from a small misalignment of the aperture with the microchannel plate detector is minimal.

\begin{figure}
\begin{center}
\includegraphics[scale=0.42]{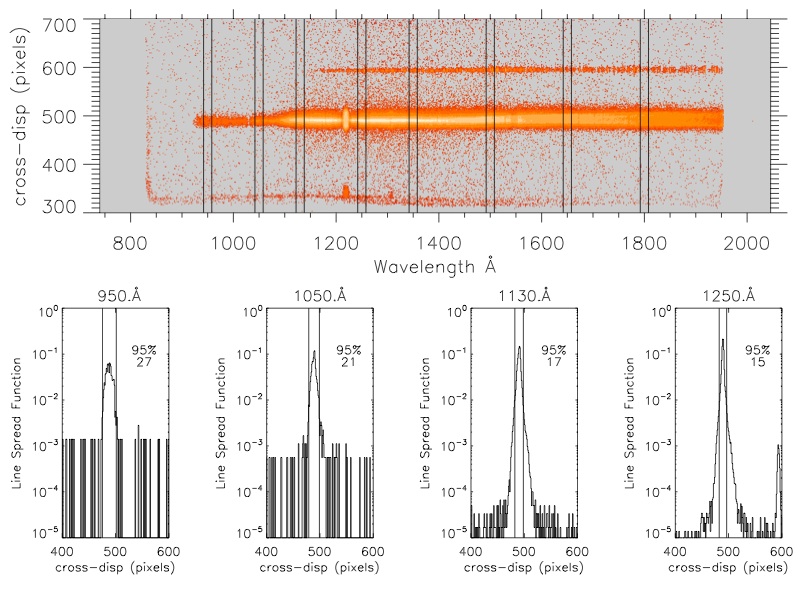}
\includegraphics[scale=0.42]{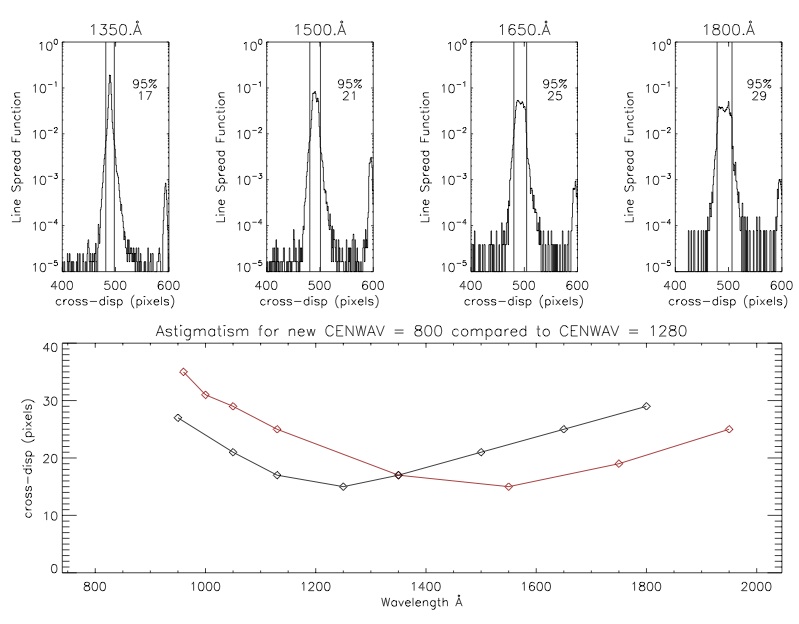}
\caption{\label{astigs1}(Top) A geometrically corrected data image with vertical lines delineating windows for calculating astigmatisms at various wavelengths.  (Middle) These astigmatism window widths within 100 \AA of the center wavelength, defined as containing 95$\%$ of the on-target counts in pixel space.  (Bottom) We plot these 95$\%$ pixel-widths (black) with the equivalent calculation for the G140L mode CENWAV=1280 (red).}
\end{center}
\end{figure}

We can compare the astigmatism in this new mode with those for existing G140L CENWAV settings in order to track the astigmatism improvements of the new CENWAV=800 mode in the FUV.  In Figure~\ref{astigs1} we define astigmatisms as extraction window widths that contain 95$\%$ of the on-target counts in each column in the geometrically corrected data, and plot them as a function of wavelength.  This figure shows a minimum width of these extraction windows at a shorter wavelength ($\sim$1250 \AA) for CENWAV=800 than that for CENWAV=1280 ($\sim$1550 \AA).

\section{Calibration}

	Turning a raw measured spectrum of AV 243 into a wavelength solution and effective area curve for the new mode involves several steps.  The onboard Platinum-Neon lamp spectrum was used to develop a wavelength calibration across the bandpass above $\sim$ 1200 \AA.  Wavelength positions were determined by calculating centroids on individual emission lines in the one-dimensional spectrum.  We found that a second order calibration solution was suitable, along with a single-frequency sinusoidal offset.  With this solution, residuals of the centroid positions to the documented PtNe lamp emission lines were limited to less than 0.1 \AA.  

\begin{figure}
\begin{center}
\includegraphics[scale=0.45]{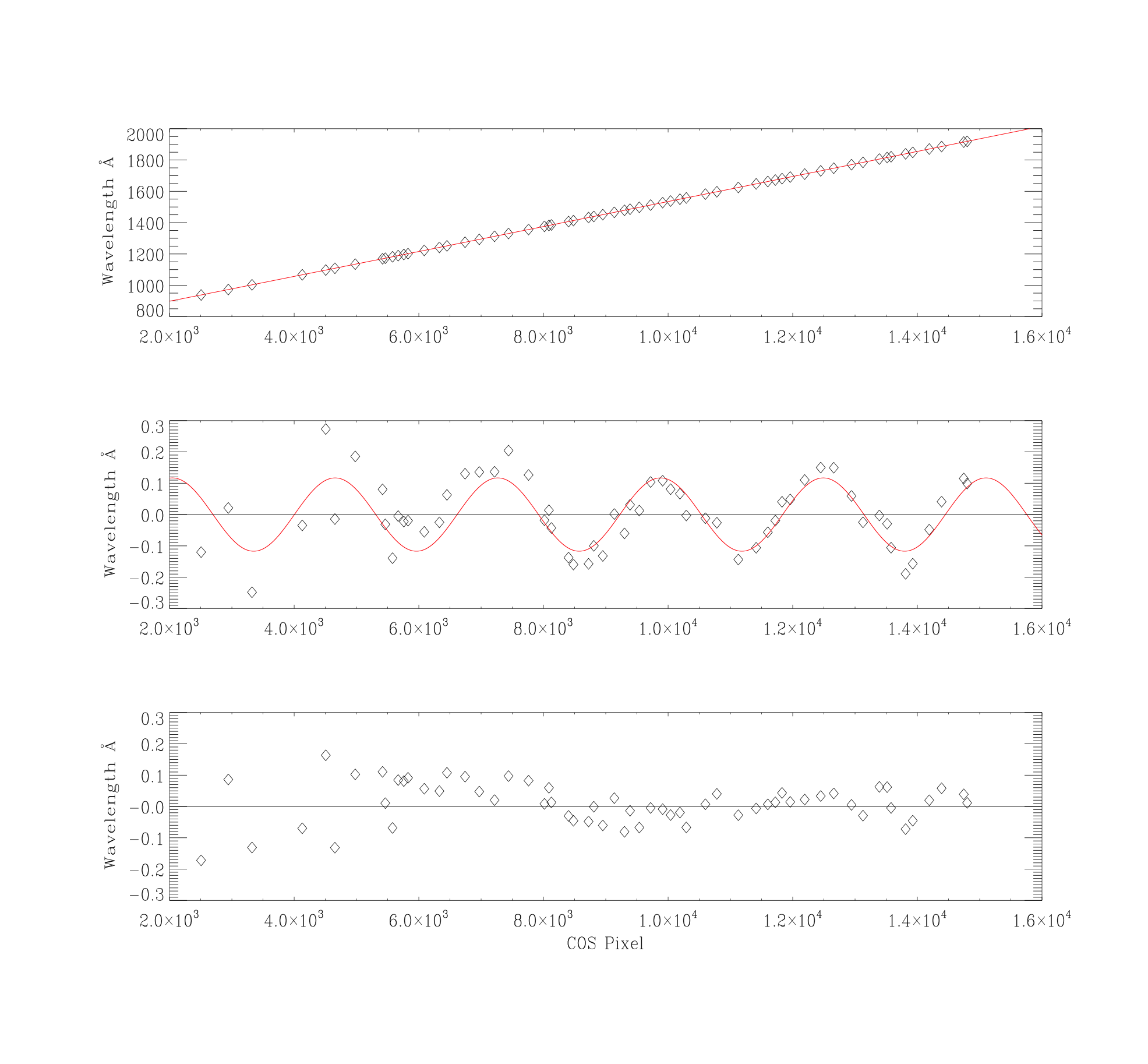}
\caption{\label{wavcal_res} (Top) The COS pixel to wavelength relationship for sixty absorption lines is plotted explicitly, with the second-order polynomial fit overplotted in red.  (Center) The same sixty points, but in this case the polynomial values at the data abscissa are subtracted from the measured wavelengths, and the residuals from this polynomial fit are plotted on the same wavelength scale.  The sinusoidal dependence of these residuals is clear, and the fit to account for this is again overplotted in red.  (Bottom) Once again the same sixty points with both the polynomial and the sinusoid fit subtracted from the measured wavelength values for each of the sixty selected pixels.  Errors are limited below 0.2 \AA.}
\end{center}
\end{figure}

Using this wavelength calibration as a starting point, we then matched the FUSE and FOS spectra of AV 243 to a theoretical absorption model of continuum emission from the star.  We found strong populations of absorbing material responsible for significant absorption features in the spectrum.  These included \ion{H}{1}, H$_2$, \ion{Ar}{1}, \ion{N}{1}, \ion{N}{2}, \ion{Si}{2}, \ion{Fe}{2}, and \ion{Fe}{3}.  The species present are typical of sight lines through the SMC including both warm and cold gas. We fitted the absorption lines, producing a model spectrum with absorption complexes (modeled as single Gaussians for our purposes) centered at $v_{helio} = +20$ and +140 km/s, consistent with expectations for the Milky Way and SMC. We compared the theoretical model with the observed COS G140L spectrum to derive the wavelength calibration at wavelengths below 1200 \AA. We considered individual features in the G140L spectrum and performed centroid calculations similar to those used for the lamp spectra at longer wavelengths.  With these offsets, we were able to fit the remaining short-wavelength region of the wavelength domain, from $\sim$950 \AA\, to $\sim$1800 \AA.  This resulting calibration function matched the observed spectrum with residuals less than 0.2 \AA.  In Figure~\ref{wavcal_res} we plot the mapping from pixel to wavelength with the quadratic solution and the sinusoid, with residuals for each step.  The final wavelength calibration parameters are shown in Table~\ref{CellParamTable}, corresponding to the pixel-to-wave mapping in the form:
\begin{displaymath}
\lambda = A+Bp + Cp^2 + D \sin\left(\omega p - \phi \right)
\end{displaymath}

\begin{figure}
\begin{center}
\includegraphics[scale=0.45]{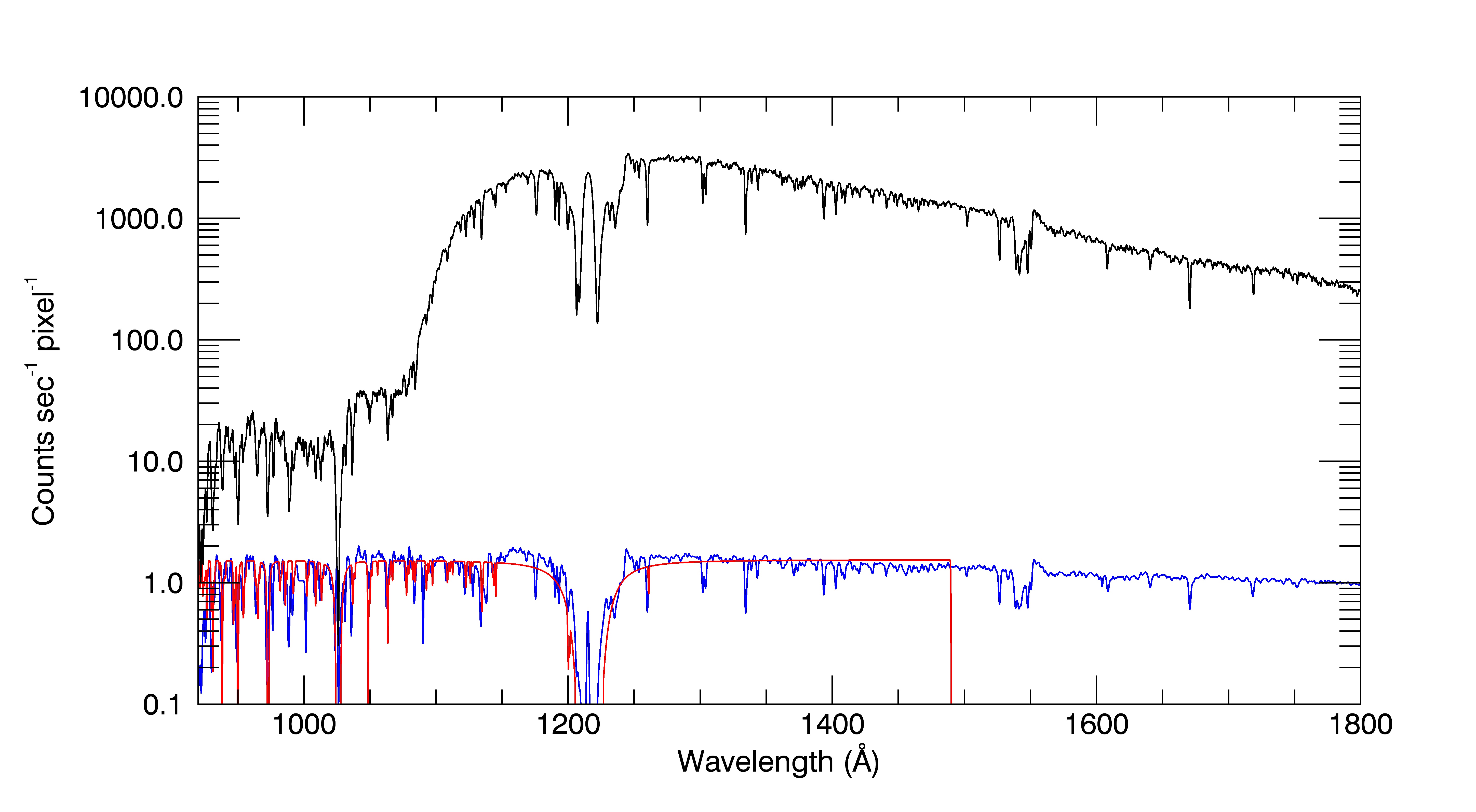}
\caption{\label{spec_oplots} Wavelength calibrated COS spectrum (black) with the FUSE/FOS reference (blue) and theoretical absorption (red) spectra.}
\end{center}
\end{figure}

\noindent
This calibrated COS spectrum with the FUSE/FOS reference and the theoretical absorption spectra can be seen in Figure~\ref{spec_oplots}.

\begin{figure}
\begin{center}
\includegraphics[scale=0.45]{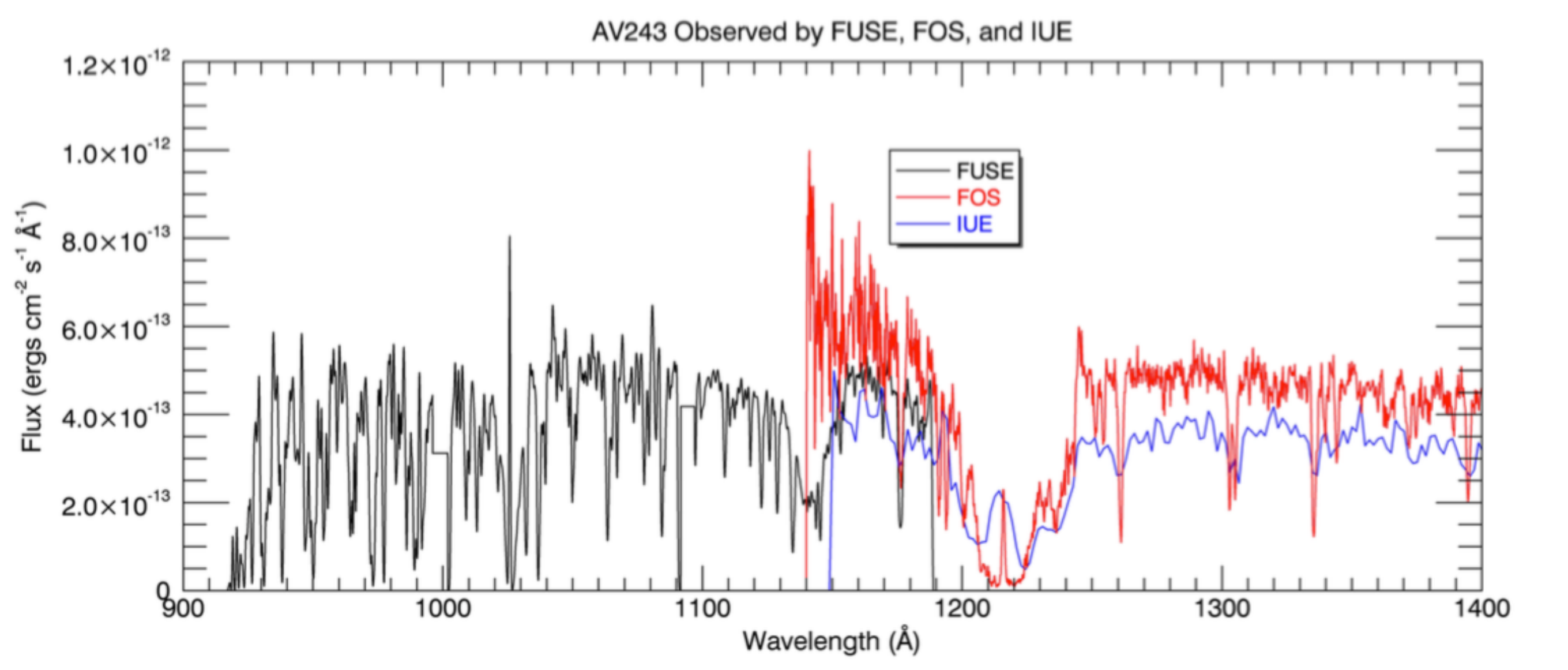}
\caption{\label{refspec_o} Spectra of the target AV 243 by FUSE, FOS, and IUE.}
\end{center}
\end{figure}

\begin{figure}
\begin{center}
\includegraphics[scale=0.45]{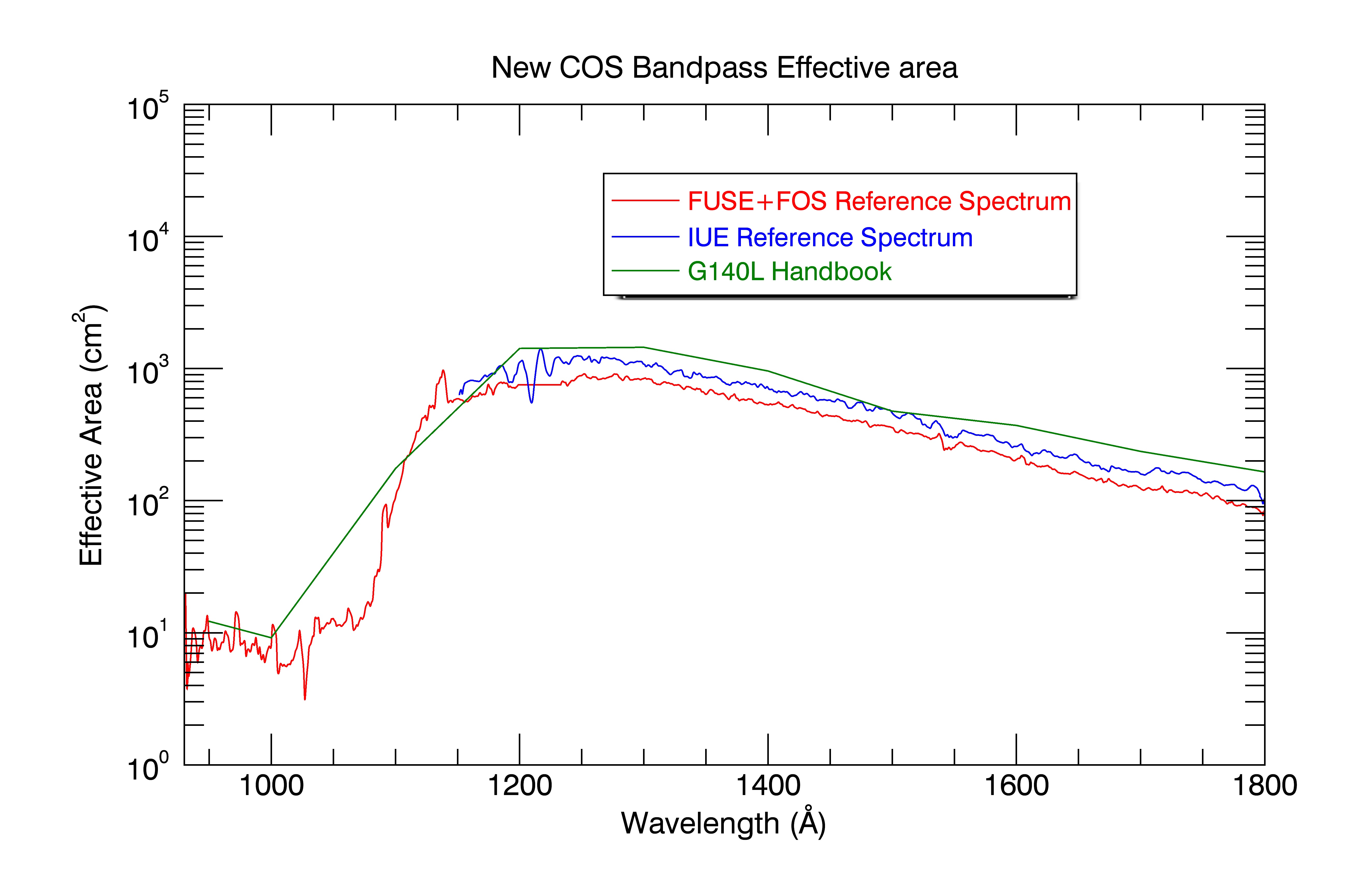}
\caption{\label{effarea2} Effective areas for a FUSE/FOS reference spectrum, an IUE reference spectrum, and the .}
\end{center}
\end{figure}

After linearizing the spectrum using this function, the total effective area of the new COS mode can be calculated by simply dividing the two spectra at any given wavelength.  We used the reference spectra previously described of the target.  We found it varied slightly between different instruments.  Figure~\ref{refspec_o} shows these archival data of the calibration target, with each providing a different effective area curve.  Figure~\ref{effarea2} provides the final effective area calculations for reference spectra of FUSE/FOS and IUE data, along with the G140L effective area from the COS handbook for when the observation was made.

\section{Results}
	Through this calibration effort, we have achieved a successful wavelength and flux calibration.  The wavelength solution is consistent with similar solutions for other modes, and suitably maps pixel space to wavelength space with errors of less than 0.2 \AA, which is a factor of $\sim$3 less than the minimum resolution limit of the G140L grating~\citep{COShandbookcycle19},  $R_{min} = \frac{\lambda}{\Delta \lambda} = 1500$.  The resolution, $\Delta \lambda$ (FWHM) at the minimum useable wavelengths, $\lambda \approx 900$ \AA, is therefore $\Delta \lambda \approx 0.6$ \AA. 
	
\begin{figure}
\centerline{
\mbox{\includegraphics[width=60mm,angle=0]{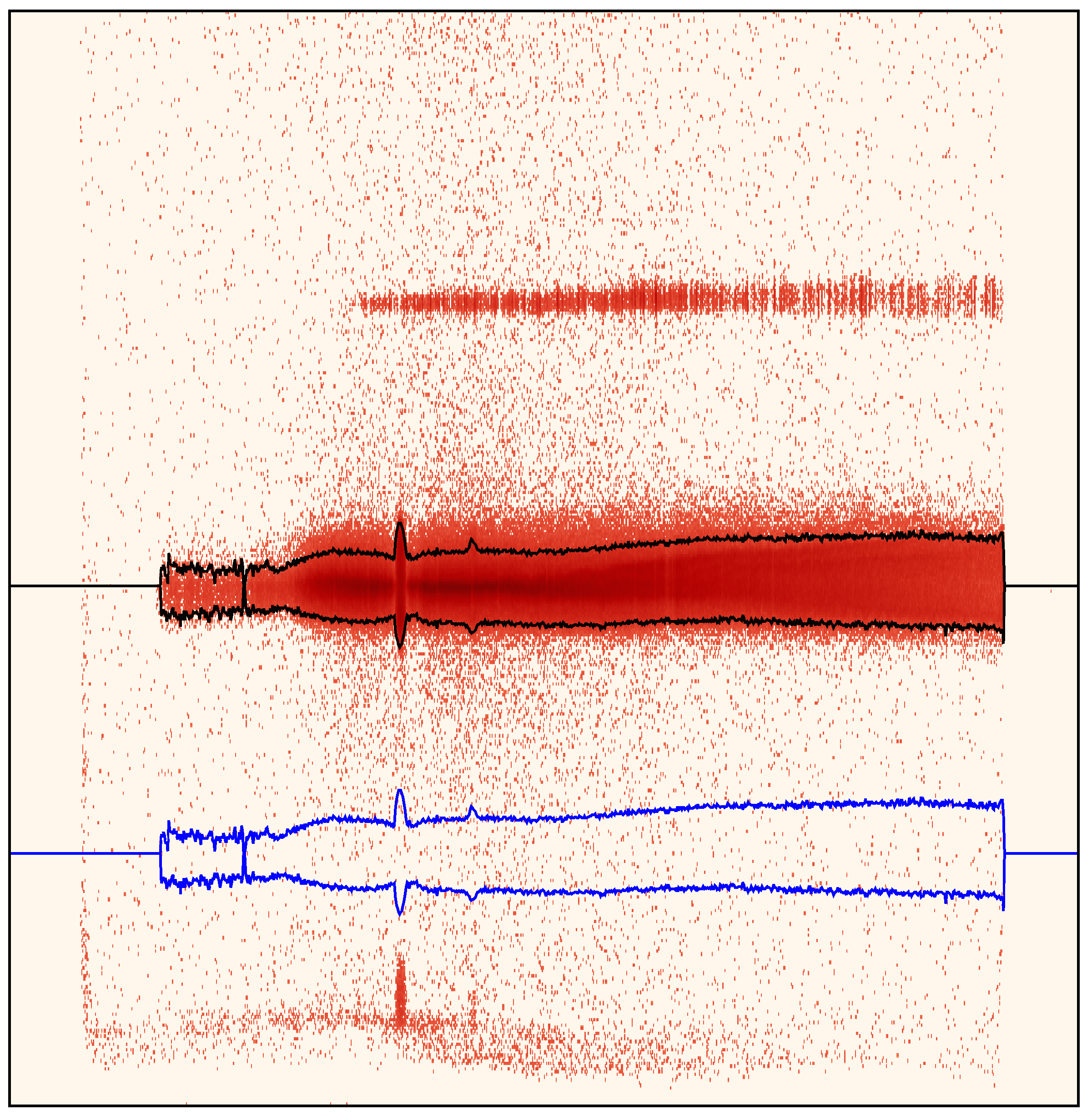}}
\mbox{\includegraphics[width=100mm,angle=0]{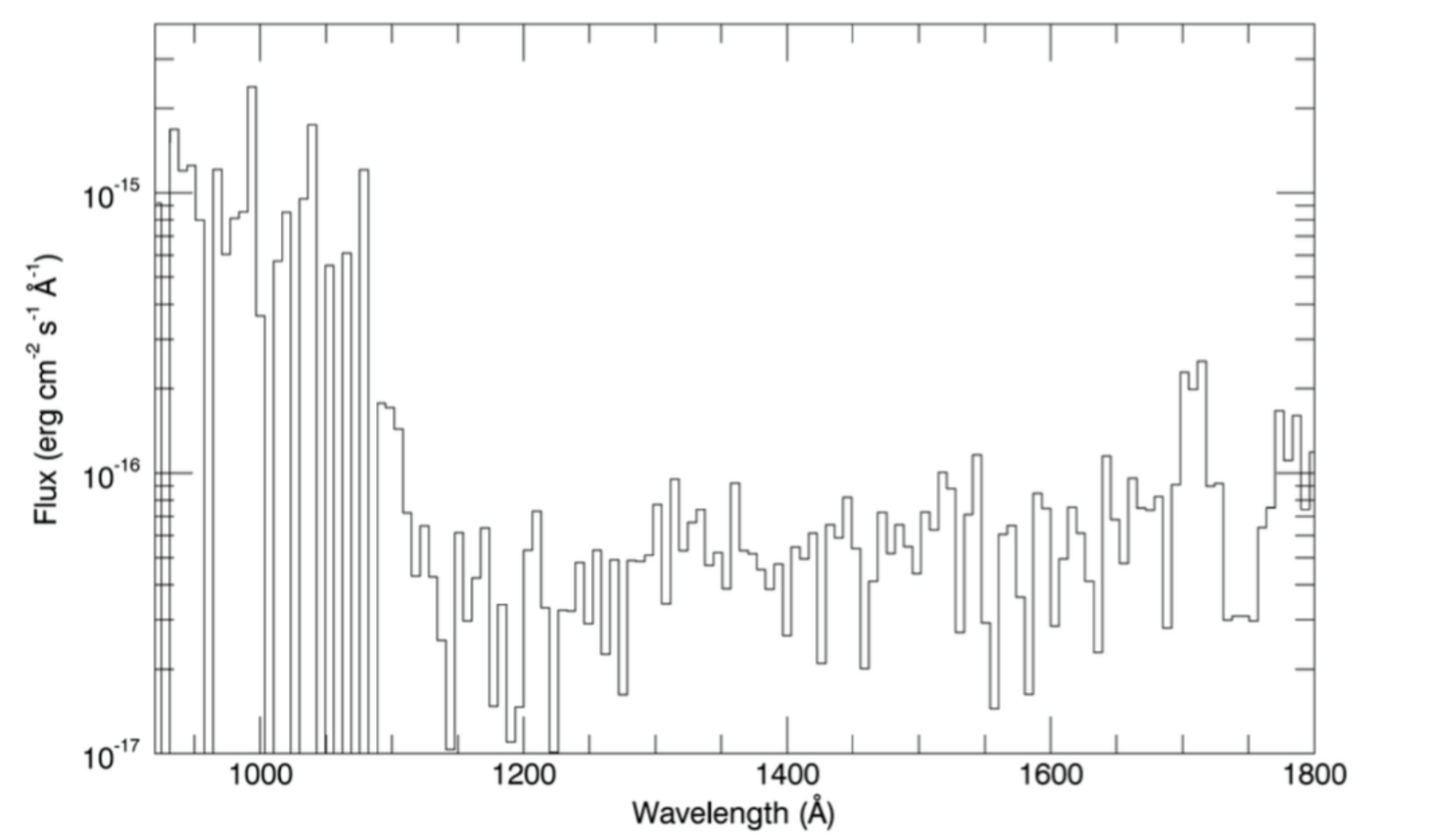}}}
\caption{\label{backg_f} (Left) Geometric corrected COS image with extraction window drawn in black around the target flux.  This same extraction window is moved to an off-target region of the image, drawn in blue, to define the background flux for the measurement.  (Right)  The one-dimensional background equivalent flux.}
\end{figure}

\begin{figure}
\begin{center}
\includegraphics[scale=0.45]{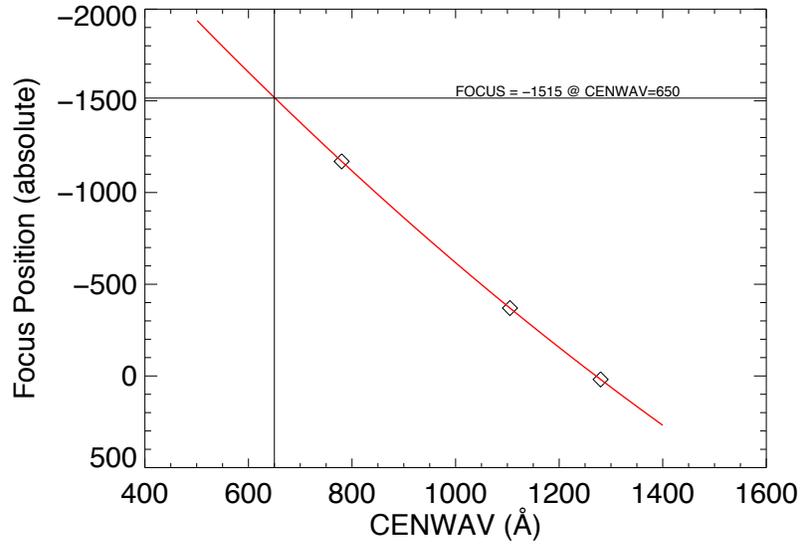}
\caption{\label{650foc} Points showing the relative focus position for CENWAV=1280, 1105, and 800, with an extrapolated curve estimating the focus positon for a potential CENWAV=650 setting.}
\end{center}
\end{figure}

	Figure~\ref{backg_f} shows a background equivalent flux for the extraction window used to define the on-target region of the detector and calculate the one-dimensional spectrum.  The extraction region is moved to an off-target region of the image, and a new spectrum is calculated from the background counts alone.  Our observations have successfully shifted the minimal astigmatism from 1550 \AA\, to 1250 \AA, suggesting that pushing further into the far-UV is possible, and a CENWAV=650 mode could be a worthwhile project in the future.  In Figure~\ref{650foc} the relative focus position of this potential mode is calculated by extrapolating a curve of focus positions for existing CENWAV modes.

\section{Discussion}
This new mode offers significant improvements over the existing modes used to observe spectra in the far-ultraviolet range.  Currently there are two observing modes for the COS G140L covering wavelength ranges from 1000 \AA\, to 2400 \AA, but the ability to observe this range is spread across two observing modes and detector segments, making it difficult to obtain science data without allowing for extra observation time and decreased data quality~\citep{COShandbookcycle19}.  The new mode enables observation between the two segments, allowing for easier and faster calibrations of flux and wavelength.  By centering this mode so that the low-astigmatism portion of the bandpass is in the 900-1100 \AA\, region, the sensitivity at these wavelengths is increased by lowering the background limit.  Additionally, looking at a complete range from 900-1850 \AA\, will no longer require a grating change, increasing potential science time and thus the total sensitivity of measurements.

There are several immediate applications for this new mode that will explore new physics.  The dynamics of young star-forming galaxies are still being explored, and Lyman-$\alpha$ photons from these sources would fall directly into the range of the new mode at a sensitivity unseen beforehand with the G140L grating.  Low redshift observations of these galaxies would greatly help in understanding how the re-ionization of \ion{H}{1} occurs at z $\sim$ 6, as it is postulated that the Lyman continuum is responsible for this ionization.  The Lyman alpha radiation could be a proxy for the Lyman continuum~\citep{Stiavelli:2004}, and observation by COS into this wavelength range will significantly aid our understanding of this phenomenon.
	
This new mode could also be useful in exploring the the \ion{He}{2} Ly$\alpha$ forest.  FUSE observations of \ion{He}{2} in the intergalactic medium (IGM) on quasar lines-of-sight (HS 1700+6416 and HE 2347-4342) showed that source hardness is related to small fluctuations in large-scale ionizing background ~\citep{Kriss:2001,Shull:2004,Zheng:2004,Fechner:2006}, and similar studies were done with G130M and G140L gratings on COS~\citep{Shull:2010,Syphers:2011,Syphers:2013}.  The new gapless COS mode opens up possibilities to put constraints on the variance of this effect in different regions of the IGM and at the redshift associated with \ion{He}{2} re-ionization.  Opening up this new mode in COS could bring background levels down to a level where \ion{He}{2} absorptions could be measured at redshifts of $\sim$2.

Another application of the new mode is the measurement of the bulk abundances of exo-planetesimals in evolved planetary systems arond white dwarfs. Planetary bodies entering within $\simeq1 R_\odot$ of the white dwarf will be tidally disrupted~\citep{Jura:2003} and subsequently accreted into the white dwarf photosphere.  Far-ultraviolet spectroscopy provides access to quantitative abundance measurements of well over a dozen elements. In hotter white dwarfs, some high-ionisation species are only available at the shorter wavelengths enabled by this mode,  (e.g. O VI 1031.9)~\citep{Barstow:2014}.
	
	Support for this work was provided by NASA through grant number 12501 from the Space Telescope Science Institute, which is operated by AURA, Inc., under NASA contract NAS 5-26555.  Boris T. G{\"a}ensicke was supported by ERC Grant Agreement n. 320964 (WDTracer).

\bibliography{master}

\begin{thebibliography}{}
\expandafter\ifx\csname natexlab\endcsname\relax\def\natexlab#1{#1}\fi

\bibitem[{{Barstow} {et~al.}(2014){Barstow}, {Barstow}, {Casewell}, {Holberg},
  \& {Hubeny}}]{Barstow:2014}
{Barstow}, M.~A., {Barstow}, J.~K., {Casewell}, S.~L., {Holberg}, J.~B., \&
  {Hubeny}, I. 2014, \mnras, 440, 1607

\bibitem[{{Boggess} {et~al.}(1978){Boggess}, {Carr}, {Evans}, {Fischel},
  {Freeman}, {Fuechsel}, {Klinglesmith}, {Krueger}, {Longanecker}, \&
  {Moore}}]{IUE_Boggess:1978}
{Boggess}, A., {Carr}, F.~A., {Evans}, D.~C., {et~al.} 1978, \nat, 275, 372

\bibitem[{{Dixon} \& {et al.}(2010)}]{COShandbookcycle19}
{Dixon}, W.~V., \& {et al.} 2010, Cosmic Origins Spectrograph Instrument
  Handbook, Version 3.0

\bibitem[{{Fechner} {et~al.}(2006){Fechner}, {Reimers}, {Kriss}, {Baade},
  {Blair}, {Giroux}, {Green}, {Moos}, {Morton}, {Scott}, {Shull}, {Simcoe},
  {Songaila}, \& {Zheng}}]{Fechner:2006}
{Fechner}, C., {Reimers}, D., {Kriss}, G.~A., {et~al.} 2006, \aap, 455, 91

\bibitem[{{Finkelstein} {et~al.}(2014){Finkelstein}, {Ryan}, {Papovich},
  {Dickinson}, {Song}, {Somerville}, {Ferguson}, {Salmon}, {Giavalisco},
  {Koekemoer}, {Ashby}, {Behroozi}, {Castellano}, {Dunlop}, {Faber}, {Fazio},
  {Fontana}, {Grogin}, {Hathi}, {Jaacks}, {Kocevski}, {Livermore}, {McLure},
  {Merlin}, {Mobasher}, {Newman}, {Rafelski}, {Tilvi}, \&
  {Willner}}]{Finkelstein:2014}
{Finkelstein}, S.~L., {Ryan}, Jr., R.~E., {Papovich}, C., {et~al.} 2014, ArXiv
  e-prints, arXiv:1410.5439

\bibitem[{{Green} {et~al.}(2012){Green}, {Froning}, {Osterman}, {Ebbets},
  {Heap}, {Leitherer}, {Linsky}, {Savage}, {Sembach}, {Shull}, {Siegmund},
  {Snow}, {Spencer}, {Stern}, {Stocke}, {Welsh}, {B{\'e}land}, {Burgh},
  {Danforth}, {France}, {Keeney}, {McPhate}, {Penton}, {Andrews},
  {Brownsberger}, {Morse}, \& {Wilkinson}}]{Green:2012}
{Green}, J.~C., {Froning}, C.~S., {Osterman}, S., {et~al.} 2012, \apj, 744, 60

\bibitem[{{Greenfield} {et~al.}(1991){Greenfield}, {Paresce}, {Baxter},
  {Hodge}, {Hook}, {Jakobsen}, {Jedrzejewski}, {Nota}, {Sparks}, \&
  {Towers}}]{FOS_Greenfield:1991}
{Greenfield}, P., {Paresce}, F., {Baxter}, D., {et~al.} 1991, in \procspie,
  Vol. 1494, Space Astronomical Telescopes and Instruments, ed. P.~Y. {Bely} \&
  J.~B. {Breckinridge}, 16--39

\bibitem[{Jura(2003)}]{Jura:2003}
Jura, M. 2003, The Astrophysical Journal Letters, 584, L91

\bibitem[{{Kriss} {et~al.}(2001){Kriss}, {Shull}, {Oegerle}, {Zheng},
  {Davidsen}, {Songaila}, {Tumlinson}, {Cowie}, {Deharveng}, {Friedman},
  {Giroux}, {Green}, {Hutchings}, {Jenkins}, {Kruk}, {Moos}, {Morton},
  {Sembach}, \& {Tripp}}]{Kriss:2001}
{Kriss}, G.~A., {Shull}, J.~M., {Oegerle}, W., {et~al.} 2001, Science, 293,
  1112

\bibitem[{{McCandliss} {et~al.}(2010){McCandliss}, {France}, {Osterman},
  {Green}, {McPhate}, \& {Wilkinson}}]{McCandliss:2010}
{McCandliss}, S.~R., {France}, K., {Osterman}, S., {et~al.} 2010, \apjl, 709,
  L183

\bibitem[{{Moos} {et~al.}(2000){Moos}, {Cash}, {Cowie}, {Davidsen}, {Dupree},
  {Feldman}, {Friedman}, {Green}, {Green}, {Gry}, {Hutchings}, {Jenkins},
  {Linsky}, {Malina}, {Michalitsianos}, {Savage}, {Shull}, {Siegmund}, {Snow},
  {Sonneborn}, {Vidal-Madjar}, {Willis}, {Woodgate}, {York}, {Ake},
  {Andersson}, {Andrews}, {Barkhouser}, {Bianchi}, {Blair}, {Brownsberger},
  {Cha}, {Chayer}, {Conard}, {Fullerton}, {Gaines}, {Grange}, {Gummin},
  {Hebrard}, {Kriss}, {Kruk}, {Mark}, {McCarthy}, {Morbey}, {Murowinski},
  {Murphy}, {Oegerle}, {Ohl}, {Oliveira}, {Osterman}, {Sahnow}, {Saisse},
  {Sembach}, {Weaver}, {Welsh}, {Wilkinson}, \& {Zheng}}]{FUSE_Moos:2000}
{Moos}, H.~W., {Cash}, W.~C., {Cowie}, L.~L., {et~al.} 2000, \apjl, 538, L1

\bibitem[{{Osterman} {et~al.}(2011){Osterman}, {Green}, {Froning},
  {B{\'e}land}, {Burgh}, {France}, {Penton}, {Delker}, {Ebbets}, {Sahnow},
  {Bacinski}, {Kimble}, {Andrews}, {Wilkinson}, {McPhate}, {Siegmund}, {Ake},
  {Aloisi}, {Biagetti}, {Diaz}, {Dixon}, {Friedman}, {Ghavamian}, {Goudfrooij},
  {Hartig}, {Keyes}, {Lennon}, {Massa}, {Niemi}, {Oliveira}, {Osten},
  {Proffitt}, {Smith}, \& {Soderblom}}]{Osterman:2011}
{Osterman}, S., {Green}, J., {Froning}, C., {et~al.} 2011, \apss, 335, 257

\bibitem[{{Sahnow} {et~al.}(2000){Sahnow}, {Conard}, {Barkhouser}, {Evans},
  {Friedman}, {Kruk}, {Moos}, \& {Ohl IV}}]{Sahnow:2000}
{Sahnow}, D.~J., {Conard}, S.~J., {Barkhouser}, R.~H., {et~al.} 2000, in
  {Instrumentation for UV/EUV Astronomy and Solar Missions}, Vol. 4139, 186

\bibitem[{{Shull} {et~al.}(2010){Shull}, {France}, {Danforth}, {Smith}, \&
  {Tumlinson}}]{Shull:2010}
{Shull}, J.~M., {France}, K., {Danforth}, C.~W., {Smith}, B., \& {Tumlinson},
  J. 2010, \apj, 722, 1312

\bibitem[{{Shull} {et~al.}(2004){Shull}, {Tumlinson}, {Giroux}, {Kriss}, \&
  {Reimers}}]{Shull:2004}
{Shull}, J.~M., {Tumlinson}, J., {Giroux}, M.~L., {Kriss}, G.~A., \& {Reimers},
  D. 2004, \apj, 600, 570

\bibitem[{{Stiavelli} {et~al.}(2004){Stiavelli}, {Fall}, \&
  {Panagia}}]{Stiavelli:2004}
{Stiavelli}, M., {Fall}, S.~M., \& {Panagia}, N. 2004, \apj, 600, 508

\bibitem[{{Syphers} \& {Shull}(2013)}]{Syphers:2013}
{Syphers}, D., \& {Shull}, J.~M. 2013, \apj, 765, 119

\bibitem[{{Syphers} {et~al.}(2011){Syphers}, {Anderson}, {Zheng}, {Smith},
  {Pieri}, {Kriss}, {Meiksin}, {Schneider}, {Shull}, \& {York}}]{Syphers:2011}
{Syphers}, D., {Anderson}, S.~F., {Zheng}, W., {et~al.} 2011, \apj, 742, 99

\bibitem[{{Zheng} {et~al.}(2004){Zheng}, {Kriss}, {Deharveng}, {Dixon}, {Kruk},
  {Shull}, {Giroux}, {Morton}, {Williger}, {Friedman}, \& {Moos}}]{Zheng:2004}
{Zheng}, W., {Kriss}, G.~A., {Deharveng}, J.-M., {et~al.} 2004, \apj, 605, 631

\end{thebibliography}

\clearpage
\begin{table}
\begin{center}
\begin{tabular}{lc}
{\bf Parameter}&{\bf Value}\\ 
\hline
A  			& 731.589			\\
B			& 0.07912			\\
C			& $3.9229\e{-8}$	\\
D  			& 0.1171			 \\
$\omega$     	& 6.274 			 \\
$\phi$		& 3.34969			\\
\hline
\vspace{0.01in}
\end{tabular}
\end{center}
\caption{Parameters of the wavelength calibration.\label{CellParamTable}}
\end{table}

\end{document}